\documentstyle[12pt]{article}
\begin{document}
\begin{center}
{\bf Is topological Skyrme Model consistent with the Standard Model?} 
\end{center}
\vspace{1in}
\begin{center}
{\bf Afsar Abbas} \\
Institute of Physics\\
Bhubaneswar-751005, India\\
e-mail: afsar@iopb.res.in\\
\end{center}
\vspace{1in}
\begin{center}
{\bf Abstract}
\end{center}
\vspace{1in}

The topological Skyrme model is known to give a successful description 
of baryons. As a consistency check, here it is shown that in view of 
the recent discovery of charge  quantization as an intrinsic and basic
property of the Standard Model and the color dependence arising therein,
the Skyrme Model is indeed completely consistent with the Standard Model.

\newpage

It is well known that in SU($N_{c}$) Quantum Chromodynamics
in the limit of $N_{c}$  going to infinity the baryons behave
as solitons in an effective meson field theory [1]. A
popular candidate for such an effective field theory is the
topological Skyrme Model [2]. It has been extensively studied for
two or more flavours [3] and it has been shown that the resemblance
of the topological soliton to the baryon in the quark model 
in the large $N_{c}$ limit is very strong [4,5]. It's baryon number 
and the fermionic character is also well understood [3,6].

Theoretically the most well studied and experimentally the
best established model of particle physics is the Standard Model
( SM ) based on the group $SU( 3_c ) \otimes SU(2)_L \otimes U(1)_Y$
[3]. The model consists 
of a priori several disparate concepts which are
brought together to give the SM its structure as a whole.
The successes of the SM are many however, it is believed to have 
a few shortcomings.It has been a folklore in particle physics that
the electric charge is not quantized in the SM. It was felt that
one has to go to the Grand Unified Theories to obtain quantization
of the electric charge. It turned out to be a false accusation against
the SM. It was clearly and convincingly demostarted in 1989/1990
that the electric charge is actually quantized in the SM [7,8].
The author showed [7,8] that the property of charge quantization 
in the SM requires the complete machinery which goes in to make it.
The SM property of having anomaly cancellation generation by 
generation, the breaking of symmetry spontaneously through
a Higgs doublet which also generates all the masses etc., all go
into bringing in quantization of the electric charge in SM.
These facts are important as there were several attempts to
demonstrate charge quantization in SM using only part
of the whole scheme, eg. using only anomaly cancelation [3].
The flaws in such logic have been pointed out by the author [8].

Also analytically the author obtained the color dependence
of the electric charge in the SM as [7] 

\begin{equation}
\newline
Q(u) = Q(c) = Q(t) = {1\over 2}(1 + {1\over N_c})
\end{equation}
\begin{equation}
\newline
Q(d) = Q(s) = Q(b) = {1\over 2}(-1 + {1\over N_c})
\end{equation}

for $N_c$ = 3 this gives the correct charges. A short derivation of
the result is given in the Appendix. It was also demonstrated by the 
author [7] that these were the correct
charges to use in studies for QCD for arbitrary $N_c$.
This was contrary to many who had been using static ( ie. independent
of color ) charges 2/3 and -1/3 [1,4,5,6].

Hence in addition to the other well known properties of the SM,
I would like to stress that the quantization of the electric charge
and the structure of the electric charge arising therein,
especially its color dependence, should be treated as an intrinsic
property of the SM. A consistency with the SM should be an essential 
requirement for phenomenological models which are supposed to
work at low energies and for any extensions of the SM which should
be relevant at high temperatures especially in the context of the
early universe.

The color dependence of the electric charge shown above should
be viewed in two independent but complementary ways.
Firstly for  $N_c \neq 3$ it is different from the static charges
Q(u)=2/3 and Q(d)=-1/3. Secondly even for $N_c = 3$ it should be
viewed as providing an anatomic view 
of the internal structure of the electric charge,
meaning as to how are 2/3 and -1/3 built up and in what way
the three colors contribute to it.
For example the SM is making the statement that in 1/3 the 3 is not
entirely due to the 3 of the QCD group $SU(3_c)$.
However this is what the SU(5) Grand Unified Theory says [9,10,11],
wherein Q(d) =-1/3 = $-1\over {N_c=3}$. This is conflict with
the SM expression where Q(d)= -1/3 = 
${1\over 2} (-1+ {1\over {N_c=3}})$.
Hence the expression for the electric charge can be a very
discriminating and restrictive tool for extensions beyond the SM.
This has been used to check consistency of various models in a 
fruitful manner [9,10,11].

Quite clearly low energy phenomenological models of hadrons
should be consistent with the SM in all respects. Is it  
true for the topological Skyrme Model? It shall be demonstrated 
below that the answer to the question in the title of the paper 
is in the affirmative.

To do so let us start with the Skyrme Lagrangian [6]
\begin{equation}
L_S = {{f_\pi}^2\over 4} Tr (L_\mu L^\mu) +
{1\over {32 e^2}} Tr {[L_\mu,L_\nu]}^2
\end{equation}

where $L_\mu = U^\dag {\partial}_\mu U$ . The U field for the
three flavour case for example is

$ U(x) = exp [{{i \lambda^a \phi^a (x)} \over {f_\pi}}]$

with $\phi ^a$ the pseudoscalar octet of $\pi$, K and $\eta$
mesons. In the full topological Skyrme this is supplemented
with a Wess-Zumino effective action

\begin{equation}
\Gamma_{WZ} = {-i\over {240 \pi^2}} \int_{\Sigma} d^5 x 
\epsilon^{\mu \nu \alpha \beta \gamma }
Tr [ L_\mu L_\nu L_\alpha L_\beta L_\gamma ]
\end{equation}

on surface $\Sigma$. Let the field U be transformed by the 
charge operator Q as 

$ U(x) \rightarrow e^{i \Lambda Q} U(x) e^{-i \Lambda Q}$.

where all the charges are counted in units of the absolute value
of the electronic charge.

Making $\Lambda = \Lambda (x)$ a local transformation the Noether
current is [6]

\begin{equation}
{J_\mu}^{em} (x) = {j_\mu}^{em} (x) + {j_\mu}^{WZ} (x)
\end{equation}
where the first one is the standard Skyrme term and the second
is the Wess-Zumino term

\begin{equation}
{j_\mu}^{WZ} (x) = { N_c \over {48 \pi^2}} \epsilon _{\mu
\nu \lambda \sigma} Tr L^\nu L^\lambda L^\sigma 
( Q + U^\dagger Q U )
\end{equation}

In the standard way [6] we take the U(1) of electromagnetism as 
a subgroup of the three flavour SU(3). Its generators can be 
found by the 
canonical methods. As the charge operator can be simultaneously
diagonalized along with the third component of isospin and
hypercharge we write it as

\[ Q =\left( \begin{array}{ccc}
q_1 & 0  & 0  \\
 0  &q_2 & 0  \\
 0  & 0  &q_3
\end{array} \right) \]

The electric charge of pseudoscalar octet mesons are known. these give

\begin{equation}
q_1 - q_2 = 1 , q_2 = q_3
\end{equation}
Hence one obtains

\begin{equation}
Q = ( q_2 + {1\over 3}) {\bf 1}_{3x3} + {1\over 2} \lambda_3 +
{1\over {2 \sqrt{3}}} \lambda_8
\end{equation}

In the standard way we use $U = A(t) U_c {(\bf x)} A(t)^{-1}$
where A is the collective coordinate.
We obtain the B=1 electric charge from the Skyrme term in terms
of the left-handed generators $L_\alpha$ only as

\begin{equation}
Q^{em} = {1\over 2} (  L_3 - {(A ^\dagger 
\lambda_3 A )}_8 {{N_c B(U_c)} \over \sqrt{3}} ) +
{1\over {2 \sqrt{3}}} ( L_8 - {(A^\dagger \lambda _8 A)}_8
{{N_c B(U_c)} \over \sqrt{3}} )
\end{equation}

The Wess-Zumino term contributes

\begin{equation}
Q^{WZ} = N_c B( U_c ) ( q_2 + {1\over 3} +
{1\over {2 \sqrt{3}}} {(A^ \dagger \lambda_3 A)}_8 +
{1\over 6} {(A^ \dagger \lambda_3 A)}_8)
\end{equation}

Hence the total electric charge is [6]

\begin{equation}
Q =  I_3 + {1\over 2} Y +
{( q_2 + {1\over 3} )} N_c B(U_c)
\end{equation}

For the hypercharge we take Y = $ N_3 \over 3 $ [12] and demanding that
the proton charge be unit for any arbitrary value of $N_c$ we find that
$q_2$ is equal to Q(d) as given in eq. (2) and hence all the correct
color dependent electric charges as demanded by the Standard Model 
are reproduced by the Skyrme model. Hence it is heartening to conclude
that the Skyrme model is fully consistent with the Standard Model.

\vspace{1in}
{\bf Acknowledgement}\\

The Author would like to thank Dr. Hans Walliser (Siegen)
for pointing out that the proper color dependent hypercharge along 
with the requirement of unit color-independent charge for the proton
be used to obtain correct charges in the Skyrme model.

\newpage

{\bf Appendix}

\vspace{1in}

To demostrate charge quantization as an intrinsic property of the
SM the complete machinery which makes the  SM is required. As required
by the SM one has the repetitive structure for each generation
of the fermions.
Let us start by looking at the first generation of quarks and leptons
(u, d, e,$\nu$ )  and assign them to
$SU(N_{c}) \otimes SU(2)_L \otimes U(1)_Y$ representation as follows 
[7,8]. 

\begin{displaymath}
q_L = \pmatrix{u \cr d}_L, (N_{c},2,Y_q)
\end{displaymath}
\begin{displaymath} u_R; (N_{c},1,Y_u) \end{displaymath}
\begin{displaymath} d_R; (N_{c},1,Y_d) \end{displaymath}
\begin{displaymath} l_L =\pmatrix{\nu \cr e}; (1,2,Y_l)
\end{displaymath}
\begin{equation}
 e_R; (1,1,Y_e)
\end{equation}

$N_c$ = 3 corresponds to the Standard Model case.
To keep things as general as possible this brings in five unknown 
hypercharges.

Let us now define the electric charge in the most general way in
terms of the diagonal generators of $SU(2)_L \otimes U(1)_Y$ as
\begin{equation} Q'= a'I_3 + b'Y \end{equation}
\newline We can always scale the electric charge once as $Q={Q'\over
a'}$ and hence ($b={b'\over a'}$)
\begin{equation} Q = I_3 + bY \end{equation}

In the SM $ SU(N_c) $ $\otimes$ $ SU(2)_{L}$ $\otimes$
$U(1)_{Y}$ is spontaniously broken through the Higgs mechanism to the
group $ SU(N_c) $ $\otimes$ $U(1)_{em}$ . In this model the Higgs is
assumed to be doublet $ \phi $ with arbitrary hypercharge $ Y_{\phi}$.
The isospin $I_3 =- {1\over2}$ component of the
Higgs develops a nonzero vacuum expectation value $<\phi>_o$. Since we want
the $U(1)_{em}$ generator Q to be unbroken we require $Q<\phi>_o=0$. This
right away fixes b in (3) and we get
\begin{equation} Q = I_3 + ({1 \over 2Y_\phi})Y \end{equation}

Next one requires that the fermion masses arise through Yukawa coupling
and also by demanding that the triangular anomaly cancels (to ensure
renormaligability of the theory) ( see [7,8] for details);
one obtaines all the unknown hypercharge in terms of the unknown Higgs
hypercharge $Y_{\phi}$. Ultimately $ Y_{\phi} $ is cancelled out
and one obtains the correct charge quantization as follows.

\begin{displaymath}
 q_L = \pmatrix{u \cr d}_L , Y_q = {{Y_\phi} \over{N_c}},
\end{displaymath}
\begin{displaymath} Q(u) = {1\over 2} ({1+{1\over N_c}}), 
                    Q(d) = {1\over 2} ({-1+{1\over N_c}})
\end{displaymath}
\begin{displaymath} u_R, Y_u = {Y_\phi} ({1+{1\over N_c}}),
                    Q(u_R) ={1\over 2} ({1+{1\over N_c}}) 
\end{displaymath}
\begin{displaymath} d_R, Y_d = {Y_\phi} ({-1+{1\over N_c}}),
                    Q(d_R) ={1\over 2} ({-1+{1\over N_c}}) 
\end{displaymath}
\begin{displaymath} l_L = \pmatrix{\nu \cr e}, Y_l = -Y_\phi,
Q(\nu) = 0, Q(e) = -1
 \end{displaymath}
\begin{equation}
 e_R, Y_e = -2Y_\phi, Q(e_R) = -1
\end{equation}

A repetitive structure gives charges for the other generation
of fermions also [7,8].

Note that the Generalized Gell Mann Nishijima expression of the
SU(6) (flavour) quark model is consistent with the above SM 
expression ( eqn. 1 and 2 ).
One takes $B = {1\over N_c}$ in the expression
Q =$I_3$ + ( B+S+C+b+t ) with the standard values of S,C,b,t
for the quarks [7].

\newpage

{\bf References}

\vspace{1in}

1. E. Witten, Nucl. Phys. {\bf B 160} (1979) 57

2. T. H. R. Skyrme, Proc. Roy. Soc. London {\bf A 260}
(1961) 127; Nucl. Phys. {\bf 31} (1962) 556

3. R. E. Marshak, "Conceptual foundations of modern 
particle physics", World Scientific, Singapore, 1993

4. G. Karl and J. E. Paton, Phys. Rev. {\bf D 30} (1984) 238

5. S. J. Perantonis, Phys. Rev. {\bf D 37} (1988) 2687 

6. A. P. Balachandran, G. Marmo, B. S. Skagerstam and A.Stern,
"Classical topology and quantum states", World Scientific, Singapore, 1991

7. A. Abbas, Phys.Lett. {\bf B 238}, (1990) 344

8. A. Abbas, J.Phys. {\bf G 16}, (1990) L163

9. A. Abbas, Hadronic J. {\bf 15} (1992) 475

10. A. Abbas, Nuovo Cim. {\bf 106 A}, (1993) 985

11. A. Abbas, Ind. J. Phys. {\bf 67 A} (1993) 541

12. H. Walliser, Phys. Lett. {\bf B 432} (1998) 15

\end{document}